\documentclass[a4paper,11pt]{article}
\usepackage{pos}

\usepackage[symbol]{footmisc}
\usepackage[english]{babel}
\usepackage{graphicx}
\usepackage{graphics}
\usepackage{braket}
\usepackage{bbold}
\usepackage{amsmath}
\usepackage{nicefrac}
\usepackage{dcolumn}
\usepackage{bm}
\usepackage{slashed}
\usepackage{datetime}
\usepackage{mciteplus}
\usepackage{multirow}
\usepackage{bbold}
\usepackage{siunitx}
\usepackage{booktabs}
\usepackage{color, soul}
\usepackage[usenames,dvipsnames]{xcolor}
\usepackage{float}
\usepackage[utf8]{inputenc}
\usepackage[normalem]{ulem}
\usepackage{mathtools}
\usepackage{setspace}

\renewcommand{\thefootnote}{\fnsymbol{footnote}}

\newcommand{\LQCD}{\Lambda_{\rm QCD}}

\newcommand{\Jps}{J/\psi}

\newcommand{\ecs}{\eta_c}

\newcommand{{\HFNRevo}}{\tt HF-NRevo}

\title{Quarkonium fragmentation in a variable-flavor number scheme: Towards NRFF1.0}
\ShortTitle{Towards NRFF1.0}

\author*[a]{Francesco Giovanni Celiberto}

\affiliation[a]{Universidad de Alcalá (UAH), Departamento de Física y Matemáticas, Campus Universitario, \\ Alcalá de Henares, E-28805, Madrid, Spain}

\abstract{We report progress on the determination and study of quarkonium production within the fragmentation approximation.
Our analyses address the moderate and large transverse-mo\-men\-tum regime, where the collinear fragmentation of a single parton is expected to dominate over the short-distance production, directly from the hard scattering, of the constituent $(Q \bar{Q})$ system.
Parton fragmentation channels to pseudoscalar and vector quarkonia are built on the basis of non-Relativistic QCD next-to-leading computations, which we use to model initial-scale fragmentation inputs. Thus, a preliminary family of Variable-Flavor Number-Scheme (VFNS) fragmentation functions, named {\tt NRFF1.0}, are constructed through standard DGLAP evolution.
Statistical uncertainties are obtained from a Monte Carlo, replica-like approach embodying missing higher-order uncertainties.}

\FullConference{31st International Workshop on Deep Inelastic Scattering (DIS2024)\\
 8–12 April 2024\\
Grenoble, France}

\begin{document}
\maketitle

\section{Opening remarks}
\label{sec:introduction}

The study of formation mechanisms of quarkonia, the so-called ``hydrogen atoms'' of QCD~\cite{Pineda:2011dg}, stands as a valuable tool to unveil core aspects of the strong force.
Quarkonium physics bridges precision studies of perturbative QCD and explorations of the proton structure. 
Hadronic decays of $S$-wave bottomonia allow for precise determinations of $\alpha_s$~\cite{Brambilla:2007cz,Proceedings:2019pra}. 
Forward emissions of quarkonia test the positivity of gluon parton densities (PDFs) at small-$x$ and -$Q^2$~\cite{Altarelli:1998gn,Candido:2020yat,Collins:2021vke,Candido:2023ujx}. 
Quarkonia provide valuable insights for 3D tomographic imaging of the proton at small~\cite{Hentschinski:2020yfm,Celiberto:2018muu,Bolognino:2018rhb,Bolognino:2021niq,Celiberto:2019slj,Silvetti:2022hyc,Kang:2023doo} and moderate $x$~\cite{Boer:2015pni,Lansberg:2017dzg,DAlesio:2020eqo,Bacchetta:2020vty,Bacchetta:2024fci,Chakrabarti:2023djs,Celiberto:2021zww}. 
Unresolved photoproductions of $\Jps$ plus a charm-jet at the EIC will ``measure'' the intrinsic-charm valence PDF in the proton~\cite{Flore:2020jau,Ball:2022qks,Guzzi:2022rca,NNPDF:2023tyk}. 
The theoretical description of quarkonium hadron\-ization is challenging. 
Numerous models have been proposed, but none fully account for all experimental observations.
To unravel the quarkonium puzzle, an effective theory, known as Non-Relativistic QCD (NRQCD), was built~\cite{Caswell:1985ui,Bodwin:1994jh}.   
NRQCD prescribes that all possible Fock levels participate to the physical-quarkonium state.
They are organized into a double series of powers of $\alpha_s$ and $v$, the latter being the $(Q \bar{Q})$ relative velocity.
NRQCD cross sections are cast as a sum of perturbative Short-Distance Coefficients (SDCs), each of them being multiplied by a nonperturbative Long-Distance Matrix Element (LDME).
For analogies with FFs for open heavy-flavored particles, see~\cite{Mele:1990cw,Cacciari:1993mq,Cacciari:2024kaa}.
NRQCD allows for rigorous testing of quarkonium production mechanisms, especially the \emph{short-distance} creation of a $(Q \bar Q)$ pair in hard scatterings, which prevails at low $|\vec p_T|$. As $|\vec p_T|$ increases, another mechanism competes: the \emph{fragmentation} of a single parton followed by its inclusive decay into the observed quarkonium~\cite{Cacciari:1994dr}. 
We describe collinear fragmentation to pseudoscalar and vector quarkonia in color singlet, via a preliminary version of our {\tt NRFF1.0} FF sets.
They builds on a new scheme, the Heavy-Flavor Non-Relativistic evolution ({\HFNRevo})~\cite{Celiberto:2024mex_article}, which makes use of NLO NRQCD initial-scale inputs and embodies a consistent DGLAP evolution and a MHOU-driven uncertainty analysis from a MC, replica-like treatment~\cite{Forte:2002fg}. 

\section{Quarkonium fragmentation from {\HFNRevo}}
\label{sec:HFNrevo}

\begin{figure*}[!t]
\centering

   \includegraphics[scale=0.37,clip]{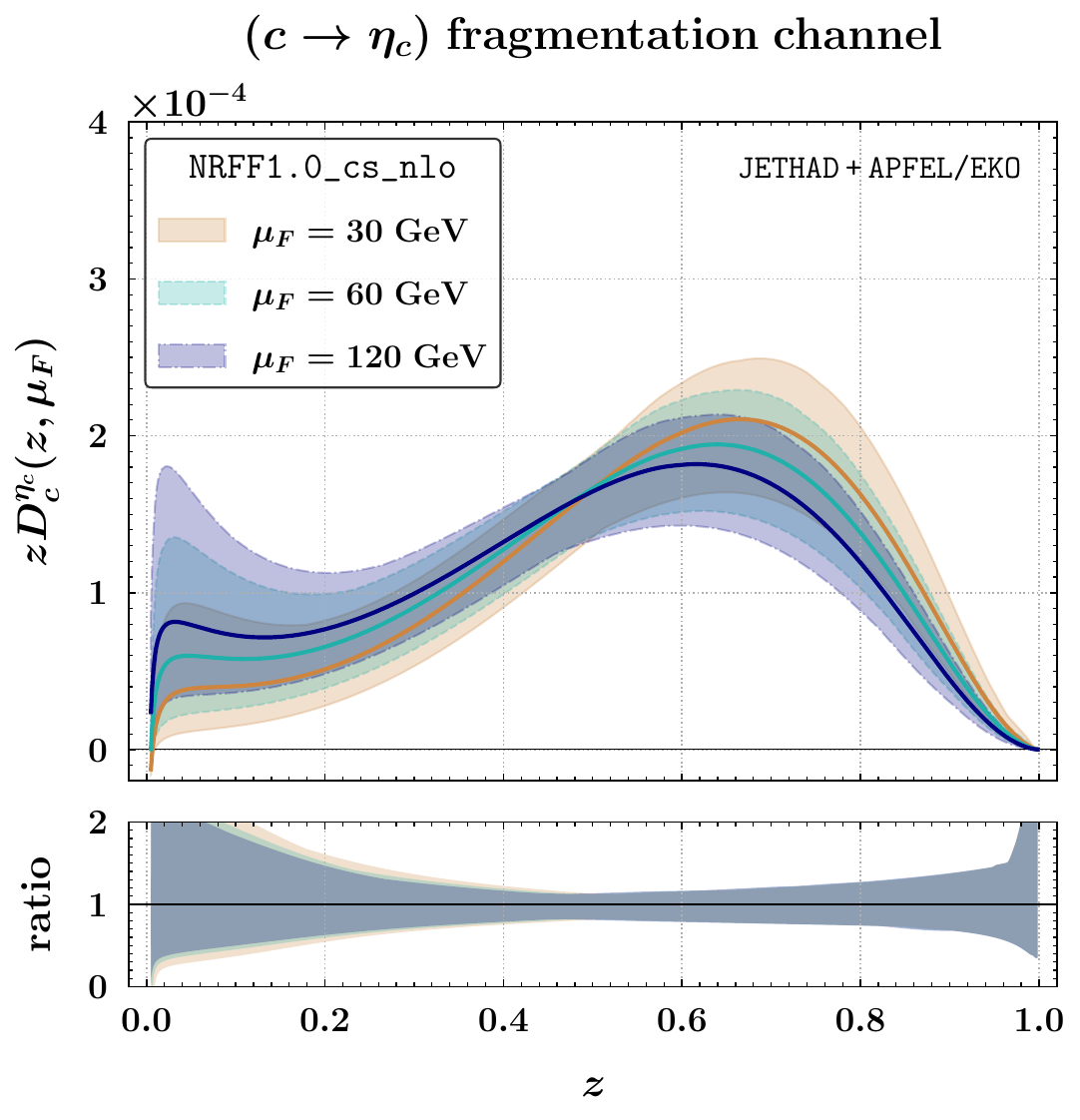}
   \hspace{0.75cm}
   \includegraphics[scale=0.37,clip]{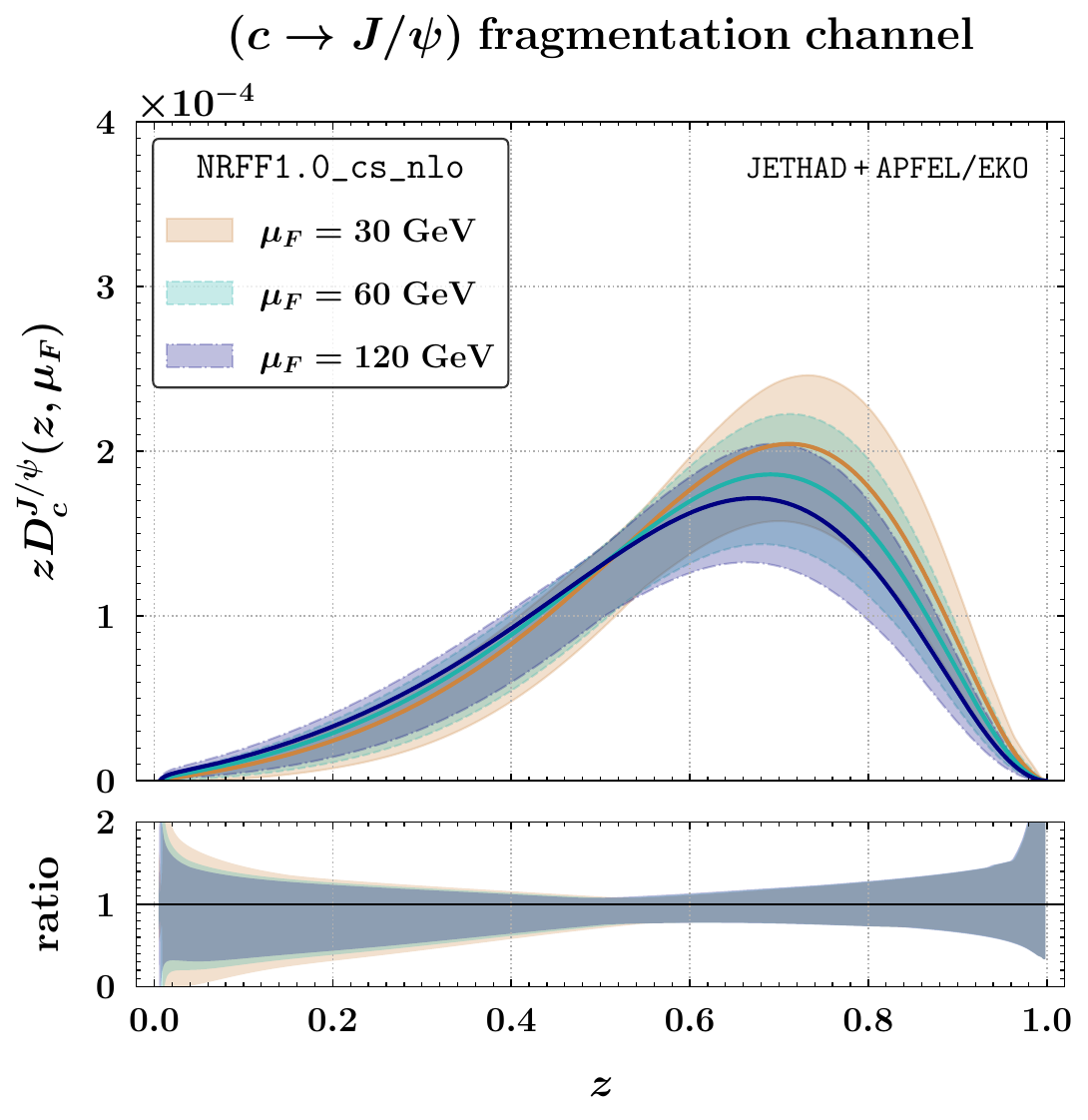}

\caption{NLO charm to color-singlet $\ecs$ and $\Jps$ FFs. Preliminary results of {\tt NRFF1.0} sets.
}

\label{fig:FFs}
\end{figure*}

Being masses of constituent heavy quarks well above $\LQCD$, initial-scale inputs of quarkonium FFs are thought to contains perturbative inputs.
Therefore, a consistent use of collinear factorization is needed here.
To this extend, we propose a novel methodology, named {\HFNRevo}~\cite{Celiberto:2024mex_article}.
It bases upon three core aspects: interpretation, evolution, and uncertainties. 
The interpretation allows one to decipher the short-distance formation at low transverse momentum ($|\vec{p}_T|$) as a two-parton fragmentation in a Fixed-Flavor Number Scheme (FFNS), and enabling subsequent FFNS-to-VFNS matching~\cite{Kang:2014tta}. This is supported by the observation that, when accounting for transverse-momentum dependence is considered, distinct singularity patterns are found in the matching tails of low-$|\vec{p}_T|$ shape functions~\cite{Echevarria:2019ynx} and moderate-$|\vec{p}_T|$ FFs~\cite{Boer:2023zit}. 
According to {\HFNRevo}~\cite{Celiberto:2024mex_article}, the DGLAP evolution of quarkonium FFs happens in two steps.
First, an \emph{expanded} and \emph{decoupled} evolution ({\tt EDevo}, done symbolically via {\tt JETHAD}~\cite{Celiberto:2020wpk,Celiberto:2022rfj,Celiberto:2022kxx,Celiberto:2020tmb,Bolognino:2021mrc,Celiberto:2021dzy,Celiberto:2021fdp,Celiberto:2022zdg,Celiberto:2022gji}), accounts for thresholds of all parton species.
Then, the standard \emph{all-order} evolution ({\tt AOevo}, done numerically via~{\tt APFEL++}~\cite{Bertone:2013vaa}; connecting {\tt EKO}~\cite{Candido:2022tld} with {\tt JETHAD} is underway) activates.
Finally, the size of MHOUs due to variations of DGLAP-evolution thresholds is gauged.
This strategy is in line with analyses on PDFs that use theory-covariance-matrix approaches~\cite{Harland-Lang:2018bxd,NNPDF:2024dpb} or the {\tt MCscales} method~\cite{Kassabov:2022orn}. 
For simplicity, we show two fragmentation channels, namely charm to charmonium FFs~\cite{Braaten:1993mp,Zheng:2019dfk,Zheng:2021ylc}. 
Left and right plots of Fig.~\ref{fig:FFs} are for $(c \to \ecs)$ and $(c \to \Jps)$ {\tt NRFF1.0} FFs, with $\mu_F$ ranging from 30 to 120~GeV.

\section{Towards {\tt NRFF1.0}}
\label{sec:conclusions}

By means of the novel {\HFNRevo} methodology, we built a preliminary version of {\tt NRFF1.0} quarkonium collinear FF sets. These functions feature color-singlet initial-scale inputs from all parton channels, calculated within NRQCD at NLO. We defined a consistent DGLAP scheme to set evolution thresholds and used a Monte Carlo replica-like treatment to address uncertainties arising from missing higher-order corrections.
The {\tt NRFF1.0} FFs are set to replace the {\tt ZCW19}$^+$ and {\tt ZCFW22} determinations currently used in the study of vector quarkonia~\cite{Celiberto:2022dyf,Celiberto:2023fzz} and $B_c$ mesons~\cite{Celiberto:2022keu,Celiberto:2024omj}. They will provide essential guidance for quarkonium physics at the HL-LHC~\cite{Chapon:2020heu,Amoroso:2022eow}, the EIC~\cite{AbdulKhalek:2021gbh,Khalek:2022bzd,Abir:2023fpo}, and future lepton machines~\cite{AlexanderAryshev:2022pkx}. 
Moreover, they will serve as a benchmark for AI-based extractions~\cite{Allaire:2023fgp,Hekhorn:2024jrj_article,Hammou:2023heg,Costantini:2024xae,Hammou:2024cwu_article}.
Future endeavors will include: exploring color-octet contributions~\cite{Cho:1995vh,Cacciari:1996dg}, implementing a general-mass VFNS~\cite{Cacciari:1998it,Forte:2010ta,Guzzi:2011ew}, and extending our studies to exotic hadrons~\cite{Celiberto:2023rzw,Celiberto:2024mrq,Celiberto:2024mab}.

\section*{Acknowledgments}
\label{sec:acknowledgments}

F.G.C. is supported by the Atracci\'on de Talento Grant n. 2022-T1/TIC-24176 (Madrid, Spain).

\vspace{-0.05cm}
\begingroup
\setstretch{0.6}
\bibliographystyle{bibstyle}
\bibliography{biblography}
\endgroup

\end{document}